\documentclass[review]{elsarticle}

\usepackage{hyperref}
\usepackage{graphicx}

\begin{document}

\begin{frontmatter}

\title{Characterization of Gravitational Waves Signals Using Neural Networks}

\author[mymainaddress]{Ana Caramete}
\author[mymainaddress,mysecondaryaddress]{Andrei-Ieronim Constantinescu\corref{mycorrespondingauthor}}
\cortext[mycorrespondingauthor]{Corresponding author}
\ead{aiconstantinescut@spacescience.ro}

\author[mymainaddress]{Lauren\c{t}iu-Ioan Caramete}

\author[mymainaddress,mythirdaddress]{Traian Popescu}

\author[mymainaddress,mysecondaryaddress]{R\u{a}zvan-Alexandru Bala\c{s}ov}

\author[mymainaddress]{Daniel Felea}

\author[mymainaddress,mysecondaryaddress]{Mircea-Victor Rusu}

\author[mymainaddress]{Petru\c{t}a \c{S}tef\u{a}nescu}

\author[mymainaddress]{Ovidiu-Mircea \c{T}\^{i}n\c{t}\u{a}reanu}

\address[mymainaddress]{Institute of Space Sciences, Romania}
\address[mysecondaryaddress]{Faculty of Physics, Bucharest University, Romania}
\address[mythirdaddress]{National Institute of Materials Physics, Romania}

\begin{abstract}
Gravitational wave astronomy has been already a well-established research domain for many years.
Moreover, after the detection by LIGO/Virgo collaboration, in 2017, of the first gravitational wave signal emitted during the collision of a binary neutron star system, that was accompanied by the detection of other types of signals coming from the same event, multi-messenger astronomy has claimed its rights more assertively.
In this context, it is of great importance in a gravitational wave experiment to have a rapid mechanism of alerting about potential gravitational waves events other observatories capable to detect other types of signals (e.g. in other wavelengths) that are produce by the same event. 
In this paper, we present the first progress in the development of a neural network algorithm trained to recognize and characterize gravitational wave patterns from signal plus noise data samples. 
We have implemented two versions of the algorithm, one that classifies the gravitational wave signals into 2 classes, and another one that classifies them into 4 classes, according to the mass ratio of the emitting source. We have obtained promising results, with 100\% training and testing accuracy for the 2-class network and $\approx95\%$ for the 4-class network.
We conclude that the current version of the neural network algorithm demonstrates the ability of a well-configured and calibrated Bidirectional Long-Short Term Memory software to classify with very high accuracy and in an extremely short time gravitational wave signals, even when they are accompanied by noise. Moreover, the performance obtained with this algorithm qualifies it as a fast method of data analysis and can be used as a low-latency pipeline for gravitational wave observatories like the future LISA Mission.
\end{abstract}


\end{frontmatter}


\section{Introduction}

\begin{figure}[h!]
\centering
\includegraphics{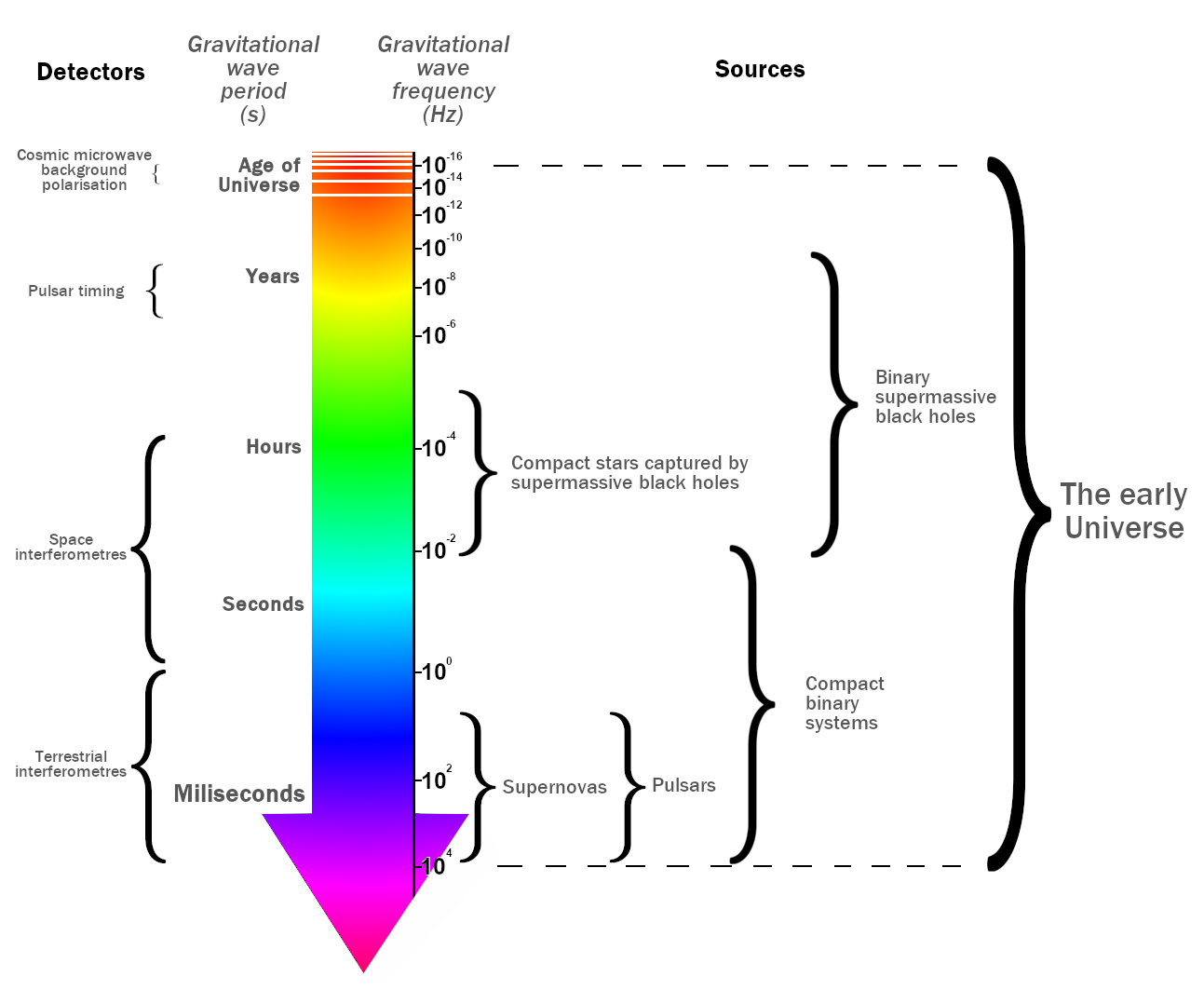}
\caption{The Gravitational Wave Spectrum, best known sources and current plus future type of gravitational wave detectors}
\label{fig:spGW}
\end{figure}

In the last decades, science has made enormous strides toward understanding the Universe. We know now that the Universe, as it stands today, started at the Big Bang and it's currently expanding at an increasing rate. By observing the electromagnetic radiation we were able to identify and characterize the initial fluctuations, that are the seeds of all cosmic structures. We have mapped the baryonic structures that form the cosmic landscape and we have found out about the existence of dark matter through its gravitational interaction with baryonic matter. Still, we don't know the answer to essential questions, such as the nature and properties of dark matter, how did the primordial black holes form in the dark matter halos, what causes the accelerated expansion and it becomes more and more clear that, in order to solve it, an extra tool, other than electromagnetic radiation, would be very useful. Since gravitation can explain many physical processes occurring in the Universe, it could be the missing messenger that carries those missing pieces of the cosmic puzzle. It is carried through the Universe by gravitational waves, perturbations in the space-time curvature that travel undisturbed from the moment of their creation. By observing and understanding the gravitational waves coming from different cosmic events we can find information about the cosmos in a new way and from a different angle and explain phenomena and processes that remain hidden to electromagnetic radiation observatories.
At this moment, we believe that most of the gravitational wave budget comes from the collisions between compact objects (such as black holes, neutron stars or neutron stars with black holes). There are also other events that generate gravitational waves, but those produced by the events listed above are much more visible than the rest, and can be used to infer new physics.
The spectrum of gravitational waves spans a wide range of frequencies and can be classified according to the emitting source. Figure \ref{fig:spGW} shows the spectrum of gravitational waves. The x-axis is the gravitational wave frequency and wavelength, on a logarithmic scale. The color code is specific to wavelengths, red = long wavelengths, blue = short wavelengths. The existing and planned type of gravitational wave detectors are placed on the left of the wavelength band, while the best known sources from which a detectable gravitational wave signal is expected are displayed on the right of the wavelength bar.

Gravitational wave astronomy has been already a well-established research domain for many years. They were proposed as a concept by Henri Poincaré in 1905, predicted by Albert Einstein in 1916, and their existence was first demonstrated after astronomers at the Arecibo Observatory in Puerto Rico discovered, in 1974, two extremely dense and heavy stars that that revolved around each other. They began to measure how the period of the stars' orbits changes over time, and eight years later they concluded that the stars approached each other at a rate identical to that predicted by general relativity. This was indirect but indisputable proof that the binary stars emitted gravitational waves. The first direct detection of a gravitational wave signal from a binary black hole merger was achieved in 2015 by the LIGO observatory \cite{Abbott2016} and was consistent with the predictions of the Einstein's theory of general relativity. Three more detections of gravitational wave signals by LIGO followed in a two year interval and, in 2017, a Nobel Prize in Physics was awarded to three core members of the collaboration. 
Currently, scientists are already planning the third generation of ground-based detectors. One of these detectors will be the Einstein Telescope, which aims to overcome current limitations related to the location of detectors by building three grouped detectors (a "multi-detector") each consisting of two interferometers that will have arms with a length of 10 km. One of interferometers will detect low frequency gravitational waves while the second will detect high frequency gravitational waves \cite{Hild2008}.

As the sensitivity of the detectors increases, more coalescence events of black holes or neutron stars, generating gravitational waves, will be detected. It will also be possible to detect new types of events such as supernovae in the local Universe. However, there are types of events that produce gravitational waves with very low frequencies and also with wavelengths that exceed the dimensions of the Earth. Such an event is impossible to detect with ground observers, because the length of the "antennas" must be large enough to detect these wavelengths. In addition, ground detectors are affected by seismic movements of the Earth and local fluctuations in the gravitational field, effects that produce a level of noise too high for the detection of very low frequency gravitational waves. Consequently, such detectors can only be constructed in space (Figure \ref{fig:spGW}) \cite{Danzmann1996,Pitkin2011}.

The Laser Interferometer Space Antenna (LISA) is one of the largest missions of the European Space Agency (ESA) to be built and launched by 2034 and it will be the first space-based gravitational wave observatory. It will consist of 3 satellites joined by laser interferometers, placed in a triangle, at a distance of 2.5 million kilometers from each other that will follow the Earth in its orbit around the Sun for an in-depth study of the Gravitational Universe.

LISA will operate in the range of very low detection frequencies, between 0.1 mHz and 1 Hz and will detect gravitational waves that have long wavelengths, produced by systems of objects with much larger masses and much wider orbits than those detected until the present. LISA will be able to detect gravitational waves produced by binary systems of ultra-compact stars in our galaxy, coalescences of supermassive black holes, binary systems of astronomical objects with very different masses that lose energy and fall into each other in a spiral-shaped trajectory ("extreme mass ration inspirals") as well as other exotic events \cite{Danzmann2017}.

A major milestone in the gravitational wave astronomy was the detection by LIGO/Virgo collaboration of the first gravitational wave signal emitted during the collision of a binary neutron star system \cite{Abbott2017} in 2017. The gravitational wave signal was accompanied by other types of signals coming from the same event over a time span ranging from 1.7 seconds to approximately 10 hours, such as the GRB 170817A observed by Fermi Gamma-ray Space Telescope and INTEGRAL, the radio observations made by Karl G. Jansky Very Large Array, optical observations made by the Las Campanas Observatory and the Hubble Space Telescope, ultraviolet observations made by the Neil Gehrels Swift Observatory or X-ray observations made by Chandra X-ray Observatory \cite{Abbott2017,Coulter2017,Soares2017,Valenti2017,Arcavi2017,Tanvir2017,Lipunov2017}.
This event is considered a major breakthrough for the field of \textbf{\textit{multi-messenger astronomy}}.
Observing the same object or event using simultaneously different types of "messengers", such as electromagnetic radiation, gravitational waves, neutrinos or cosmic rays has proven to be essential for maximizing the science harvested from the detection/observation. For example, while gravitational waves and neutrinos have the potential to reveal sources otherwise invisible to the electromagnetic and cosmic rays detectors, photons can be used in determining the localization of the source, the host galaxy and in characterizing its environment. In turn, knowing an accurate position of the source leads to a better knowledge of the intrinsic parameters, such as spin and source mass.

In this context, it is of great importance in a gravitational wave experiment to have a rapid mechanism of alerting the possible complementary observatories in order to extract the maximum information from the detected gravitational wave sources.

For observatories like the LISA Mission, two of the key ingredients to succeed in making a multi-messenger detection and also in identification and characterization of the gravitational waves will be (a) the generation of alerts of potential gravitational waves events towards other space and Earth observatories capable to detect other types of signals (e.g. in other wavelengths) that are produce by the same event, and (b) the identification of the protected periods for observations, during which gravitational wave events are more likely to be detected and the all of the maintenance or calibration activities should be suspended. All of the above will be done by the Low-Latency Pipeline system within the LISA Mission.

The set of parameters that characterize the collision of two black holes is composed of: the parameters of the orbits before the collision, the orientation of the spin of each black hole before the collision, their mass ratios and a considerable number of extra parameters, if we consider also the environment around the black holes. The time required to estimate these parameters is very long because of all the very laborious calculations that are needed to simulate various physical processes and to analyze and filter the gravitational wave signal. In order to improve the time needed for the determination of the parameters, several optimization techniques are used, such as Monte Carlo Markov techniques \cite{Raymond2009} or interpolation of waveforms \cite{Smith2013}, which have proven their effectiveness and efficiency.

However, as more and more advanced detectors appear, it is important to reduce even further the computation time for these determinations (to achieve low latency and to produce alerts), especially in the current context of complementary observations, which uses several observation facilities located on the ground or in space. Thus, in recent years, new methods have been investigated to quickly characterize gravitational wave sources, which use the analysis of signal properties (assuming it has been filtered) and strategies to treat the problem in reverse, so starting from simulating the physical phenomena and then the signal to identify different types of signals observed \cite{Carrillo2015}. The latter approach proved to be solved very efficiently by the use of machine learning techniques, more precisely by involving a neural network with simulated gravitational wave signals to recognize and characterize the observed gravitational waves.

Machine learning techniques such as artificial neural networks already have applications in various disciplines including gravitational wave astronomy for detecting and characterizing multiple signals of gravitational waves from black hole systems \cite{George2018,Carrillo2015,LeCun1998,Lin2020,Gabbard2018, Morvan2020}.

In this paper we present the first results in the development of a fast and accurate data analysis pipeline based on a neural network algorithm trained to recognize and characterize gravitational wave patterns in signal + noise data samples. The purpose of this pipeline is to create the capability for gravitational wave observatories (like the LISA Mission) to generate low-latency alerts for other space or Earth observatories that have the ability to react quickly and redirect their instruments in the direction from which gravitational wave signals are announced, with the purpose of detecting complementary signals (e.g. gamma ray burst). This would be an essential tool in the context of multi-messenger observations. 

The paper is structured into four sections. In the introductory section we describe the context in which this work was done and its purpose. In section 2 we give a description of deep neural networks in general and of bi-directional long short term memory networks in particular, followed by a detailed description of the steps taken in the development of our pipeline. Then, in section 3 we present the results obtained. In section 4 we give the conclusions.

\section{Identification and classification of gravitational wave signals using neural networks}
\label{section:main}
As already mentioned before, machine learning techniques such as artificial neural networks have already been used in gravitational wave astronomy to detect and characterize gravitational wave signals coming from binary black-hole systems.
The analysis of the gravitational wave signals using neural networks has proven one of the fastest and most accurate methods of characterization of the sources, being capable of processing a large number of parameters in a short amount of time \cite{George2018, George2018a, LeCun1998, Lin2020, Gabbard2018, Morvan2020}.

We developed a neural network based on a Bidirectional Long-Short Term Memory (BiLSTM) deep learning algorithm, hereinafter referred to as LL BiLSTM ("Low-Latency Bidirectional Long-Short Term Memory”), which we trained to recognize gravitational waveforms and classify them according to the parameters of the emitting source. For training and testing the network, we used simulated LISA-like data, made using an in-house developed code \cite{Popescu2020}.
The Neural Network is implemented using MATLAB’s \href{https://www.mathworks.com/products/deep-learning.html}{Deep Learning Toolbox™} framework.

In what follows, we will first present a general description of deep neural networks in general and BiLSTM networks in particular, and then we will present the details concerning the development of the neural network and the results obtained.

\subsection{Deep Neural Networks}

Neural networks that contain more than three layers of neurons (including the input and output) are called deep neural networks. Their training is called deep learning.

Deep learning has become popular in recent years \cite{Bahaadini2017,Gabbard2018,George2018,Li2017,Gebhard2019}, especially with the rapid development of graphics processing technology. Some of the many applications of implementing this procedure are image processing \cite{Razzano2018,Wei2019,Torres2016}, medical diagnoses \cite{Devunooru2020} and gene expression classification \cite{Tirumala2019}. Recently, they have also been used successfully for gravitational wave astronomy in the form of classification of gravitational wave signal errors \cite{Bahaadini2017} in which it was demonstrated that deep learning can be used as a detection method \cite{Gabbard2018} and estimation of source parameters. In the context of gravitational wave detection, the latest efforts have focused on detecting gravitational wave signals from binary black hole systems using convolutional neural networks \cite{Gebhard2019,George2018}.

Deep learning can make quick analyzes because a lot of the calculations are done beforehand, in the training phase. This process helps with low latency research that has the potential to be several orders of magnitude faster than other classification methods \cite{Gabbard2018}.

A deep learning algorithm can be formed by several processing layers called neurons, which can be made up of several strings of inputs. A neuron will have a filtering function that will transform the input data strings. This transformation is a linear operation between the input strings on one side and the weights and adjustment parameters associated with the neurons ("bias parameter") on the other.
The layer resulting from this operation goes through a non-linear activation function to constrain the data output in a finite interval. Deep learning algorithms consist of an input neuron layer, one or more hidden layers, and an output layer. The scales produced by the last layer of neurons correspond to the probability that the input sample belongs to a certain class, so each neuron correspond to a probability \cite{Gabbard2018}.

\paragraph{Bidirectional Long-Short Term Memory (BiLSTM) Neural Networks}

BiLSTM networks are recurrent neural networks (RNNs) which are a class of neural networks that specialize in sequential data processing (x(1),…, x(t)). Recurrent networks can scale to longer data sequences than networks that do not specialize in sequence processing can perform. They can also process sequences with variable lengths.

All recurrent neural networks have a common structure, i.e. at a certain point t takes only information from the past (x(1),…, x(t-1)) and the current input x(t). These models allow information from past data to influence the present state.

In many applications we want the output data to depend on all data, both input and intermediate. For example, in voice recognition, the correct interpretation of the current sound as a phenomenon depends on the following sounds recorded due to the co-articulations, and may also depend on the following words addressed due to the word dependencies in that language. If there are two plausible interpretations of the same word, we will look for the exact meaning in both the words in front and back in the sentence if necessary. It is applied to the recognition of writing and other automated machine learning tasks. Recurrent bidirectional neural networks have been invented precisely to address such problems.

While a normal recurrent network can only look at the time-sequences in one direction, from the beginning to end, a bidirectional neural network analyzes the sequence from both directions, reducing the errors and increasing the precision.

LSTM networks are a type of recurrent neural networks that proved to be efficient in learning sequence and time-series data. Also, an LSTM network can learn long-term dependencies between time steps of a sequence. 
The forward propagation equations for the architecture of a recurrent artificial neural network are also given below. The cells are recurrently connected to each other, thus replacing the usual hidden units of a common recurrent neural network. An input characteristic is calculated with an ordinary artificial neuron unit. Its value can be summed in a single state if the input sinusoidal signal allows this. The status unit has its own automatic loop whose weight is controlled by the forgetting gate. The cell result can be stopped by the exit gate. All bearing units have a sinusoidal non-linearity while the input units have a condensed non-linearity.

Deeper architectures have also been used successfully. Instead of a unit that applies non-linearity in the order of elements to link input transformations to recurring units, recurrent neural networks with long-term short-term memory have special cells that have an internal recurrence to overcome the limitations of a neural network common recurrences. Each cell has the same input and output data as a neural network common recurrent, but has several parameters and a system of fencing units that control the flow of information. The most important component is the base unit si(t), which has a linear automatic loop. The weight of the automatic loop is controlled by a forgetting gate unit fi(t) which sets values of weights between 0 and 1 by a sigmoidal unit:

\begin{equation} \label{eq:1}
f_{i}^{(t)} = \sigma\left(b_{i}^{f} + \sum\limits_{j} U_{i,j}^{f} x_{j}^{(t)} + \sum\limits_{j} W_{i,j}^{f} h_{j}^{(t-1)}\right)
\end{equation}

where x(t) represents the current input vector, h(t) represents the hidden layer vector, which contains the results of all cells in the LSTM, and bf, Uf and Wf are the adjustment (bias), the input weight and the recurrent gate weight, of forgetfulness gates. The LSTM cells of the internal states are calculated taking into account the condition of the automatic loop weight fi(t):

\begin{equation} \label{eq:2}
s_{i}^{(t)} = f_{i}^{(t)} s_{i}^{(t-1)} +  g_{i}^{(t)}\sigma\left(b_{i} + \sum\limits_{j} U_{i,j}^{f} x_{j}^{(t)} + \sum\limits_{j} W_{i,j}^{f} h_{j}^{(t-1)}\right)
\end{equation}

where b, U and W represent the adjustment (bias), the input weights and the recurrent weights in the LSTM cells, respectively. The external entrance gate gi(t) is calculated similarly to the forgetting gate but with its own parameters.

\begin{equation} \label{eq:3}
g_{i}^{(t)} = \sigma\left(b_{i}^{g} + \sum\limits_{j} U_{i,j}^{g} x_{j}^{(t)} + \sum\limits_{j} W_{i,j}^{g} h_{j}^{(t-1)}\right)
\end{equation}

The result hi (t) of the LSTM cell can be closed by the exit gate qi (t) which can be used as a sinusoidal unit for porting:

\begin{equation} \label{eq:4}
h_{i}^{(t)} = \tanh \left(s_{i}^{(t)} \right) q_{i}^{(t)}
\end{equation}

\begin{equation} \label{eq:5}
q_{i}^{(t)} = \sigma\left(b_{i}^{0} + \sum\limits_{j} U_{i,j}^{0} x_{j}^{(t)} + \sum\limits_{j} W_{i,j}^{0} h_{j}^{(t-1)}\right)
\end{equation}

which has the parameters b0, U0, W0 for adjustment (bias), the input weights and the recurrent weight respectively. Alternatively, we can choose the status cell and (t) as an additional input that can be weighted into three portions of the unit i. For this we need three more additional parameters.

Recurrent neural networks with long-term short-term memory can learn long-term dependencies more easily than simple recurrent neural networks, primarily for artificial databases designed to test the ability to learn long-term dependencies, as well as for other more special problems \cite{Graves2012,Graves2013,Sutskever2014}.

\subsection{LL BiLSTM Neural Network}

Based on the principles described in \cite{Gari2017} we have developed LL BiLSTM (which is short from \textit{\textbf{Low-Latency Bidirectional Long-Short Term Memory}}), a \href{https://ch.mathworks.com}{Matlab} application that embeds a Bidirectional Long-Short Term Memory (BiLSTM) deep learning algorithm trained to recognize gravitational waveforms and classify them according to  parameters of the emitting source.
So far, LL BiLSTM has been used for the classification of gravitational waveforms emitted by compact binary systems according to the mass ratio of objects in the system, \textit{q}.
There are two types of algorithms implemented in the application:
\begin{itemize}
    \item A BiLSTM algorithm that classifies waveforms according to two classes: H(\textit{\textbf{high}}), representing gravitational waves emitted by binary sources with high mass ratio, $q\in[400 -500]$, and L(\textit{\textbf{low}}) representing
    gravitational waves emitted by binary sources with low mass ratio, $q\in[1-10]$.
    \item A more complex version of the same algorithm, that classifies waveforms according to four classes, A, B, C and D, where: class A is the class of gravitational waves emitted by binary sources with mass ratio $q\in[1-10]$, class B is the class of gravitational waves emitted by binary sources with mass ratio $q\in[300-350]$, C is the class of gravitational waves emitted by binary sources with mass ratio $q\in[400-500]$ and D is the class of gravitational waves emitted by binary sources with mass ratio $q\in[750-950]$.
\end{itemize}
LL BiLSTM also contains two modules for (pre-)processing the \textit{\textbf{raw-data}} files provided by the gravitational wave simulator and producing more complex data structures that pairs each simulated waveform timeseries with two source parameters: the chirp mass corresponding to the ratio \textit{q} as well as a label corresponding to each class ("H" or "L" in the case of the two-class algorithm and "A", "B", " C” and “D” in the case of the 4-class algorithm. These complex data structures are the baseline for creating the training and testing datasets.
LL BiLSTM app employs Matlab's \href{https://www.mathworks.com/products/deep-learning.html}{\textit{\textbf{Deep Learning Toolbox™}}}. Also, in order to improve the performance and to reduce the running time of the algorithm, we have used feature extraction techniques - fast Fourier transform analysis. This required the Matlab's \href{https://www.mathworks.com/products/signal.html}{Signal Processing Toolbox™}.

\subsection{Training and Testing Datasets}
\label{subsection:trainandtestdata}
The data used to train and test the neural network consists of a set of simulated data representing gravitational wave signals (the amplitude of the gravitational wave as a function of time) emitted by a binary system of black holes orbiting their common center of mass. The simulations were performed using a custom made Matlab code \cite{Popescu2020}.
The waveforms were generated in the quadrupole approximation, non-spinning point mass approximation and circular orbits approximation. The simulation generates time series for the two gravitational waves polarizations modes, $h_{+}(t)$ and $h_{x}(t)$ for different values of the source parameters: the mass ratio of the two objects that form the binary system, \textit{q}, the orbital inclination \textit{i} and the distance to the source, \textit{r}. One of the masses of the binary system has been fixed at $10^3$ solar masses.The total polarization, $h_{tot}$, is thus a linear combination of the form $h_{tot} = a\cdot h_{+}+ b \cdot h_{x}$, where a and b are two coefficients whose values depend on the antenna pattern. For example, in the case of LIGO detector, these coefficients have values between 0 and 1.
At this stage of the study we focused mainly on understanding and generating the "theoretical" waveform, without imposing restrictions related to, for example, the characteristics of the LISA instrument, the slow-motion approximation or other constraints which would add constraints to the parameter space used in the simulation code (q, i and r).
Figures \ref{fig:difq} to \ref{fig:difr} below show how the waveforms change when varying the three source parameters, one by one. We have plotted the time evolution of the gravitational wave amplitudes, for four distinct values of the three source parameters considered, namely the mass ratio \textit{q}, the orbital inclination \textit{i} and the distance to source \textit{r}, while keeping the other two constant. Each figure illustrates the waveforms generated for the four distinct values of the parameters, both individually and superimposed.
In Figure \ref{fig:difq}, it can be seen how the frequency and amplitude of a gravitational wave emitted by a binary system increase with the mass ratio between the two objects in the system, \textit{q}. The other two source parameters are fixed to $r = 10^{22} cm$ and $i = 0 rad$. In addition, from Figure \ref{fig:difq1} it can be seen that the signal duration increases as the objects have more similar masses.
In Figure \ref{fig:difi}, it can be seen how the frequency remains constant and the amplitude increases with the orbital inclination, \textit{i}. The other two source parameters are fixed to $r = 10^{22} cm$ and $q = 550$.
In Figure \ref{fig:difr}, we can see how the frequency of the wave remains constant while the amplitude decreases when increasing the distance to the source of gravitational waves, \textit{r}. The other two parameters of the source are fixed to $i = 0 rad$ and $q = 550$.

\begin{figure}[h!]
\centering
\includegraphics[viewport=0cm 0.2cm 20cm 6cm,clip,scale=0.8]{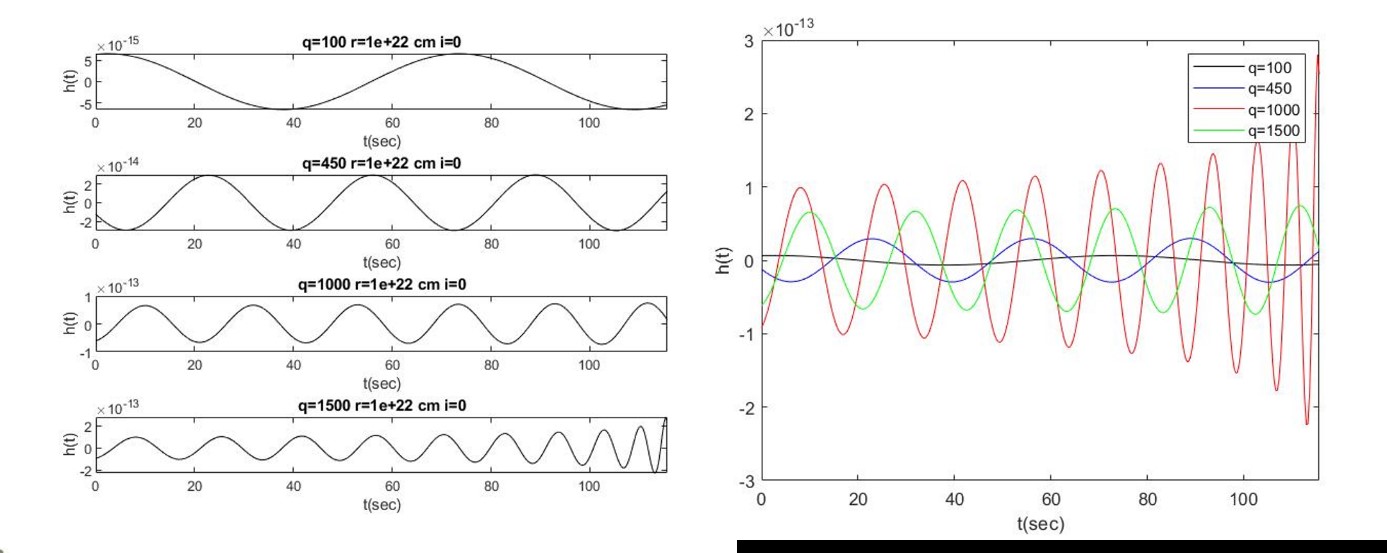}
\caption{Change in the shape of gravitational waves emitted by a binary system when varying the mass ratio of the two objects in the system. Left panel: Four individually waveforms, for four different source mass ratios. Right panel: the same four waveforms, superimposed.}
\label{fig:difq}
\end{figure}

\begin{figure}[h!]
\centering
\includegraphics[viewport=2cm 0cm 20cm 14cm,clip,scale=0.8]{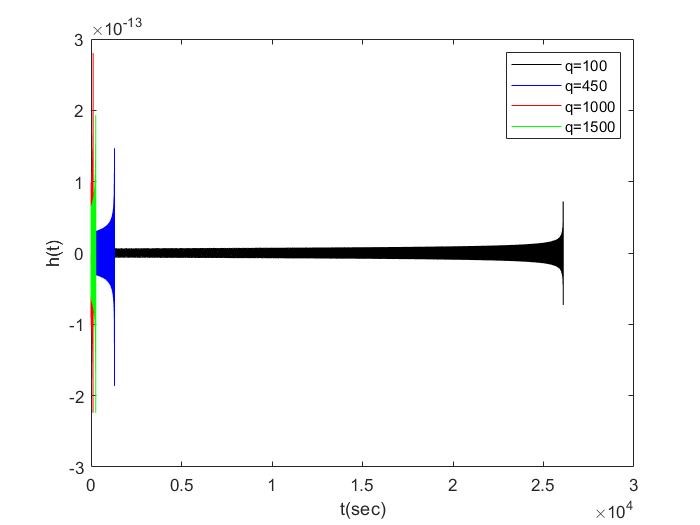}
\caption{The gravitational waves emitted by binary systems, for four different source mass ratios. The signal duration increases as the two objects forming the system have more similar masses.}
\label{fig:difq1}
\end{figure}

\begin{figure}[h!]
\centering
\includegraphics[viewport=0.5cm 0.2cm 20cm 6cm,clip,scale=0.8]{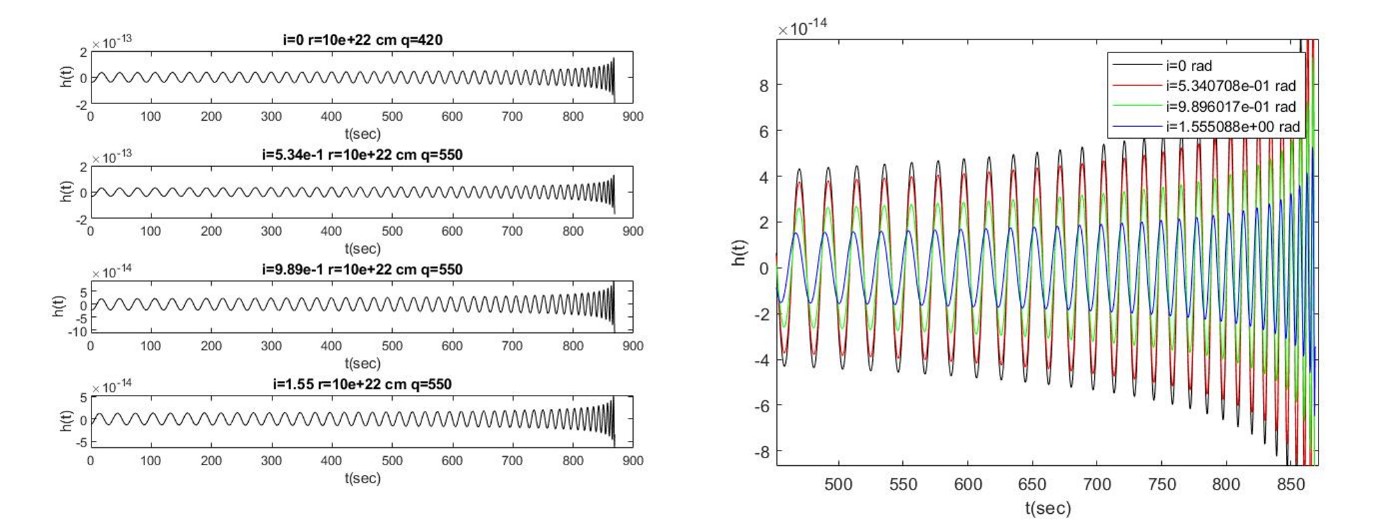}
\caption{Change in the shape of gravitational waves emitted by a binary system when varying the orbital inclination of the source. Left panel: Four individually waveforms, for four different orbital inclinations of the source. Right panel: the same four waveforms, superimposed.}
\label{fig:difi}
\end{figure}

\begin{figure}[h!]
\centering
\includegraphics[viewport=0.5cm 0.3cm 20cm 7cm,clip,scale=0.8]{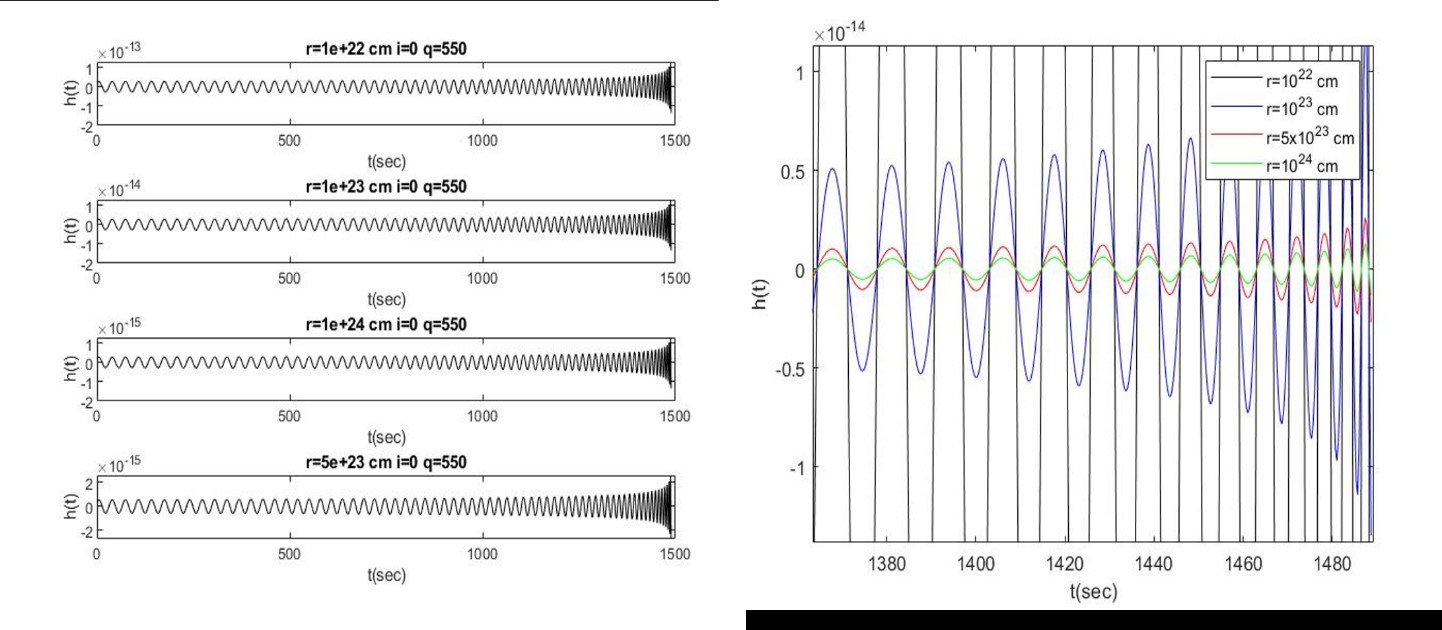}
\caption{Change in the shape of gravitational waves emitted by a binary system when varying the distance to the source. Left panel: Four individually waveforms, for four different distances. Right panel: the same four waveforms, superimposed.}
\label{fig:difr}
\end{figure}

\begin{figure}[h!]
\centering
\includegraphics[viewport=0cm 0cm 50cm 30cm,clip,scale=0.3]{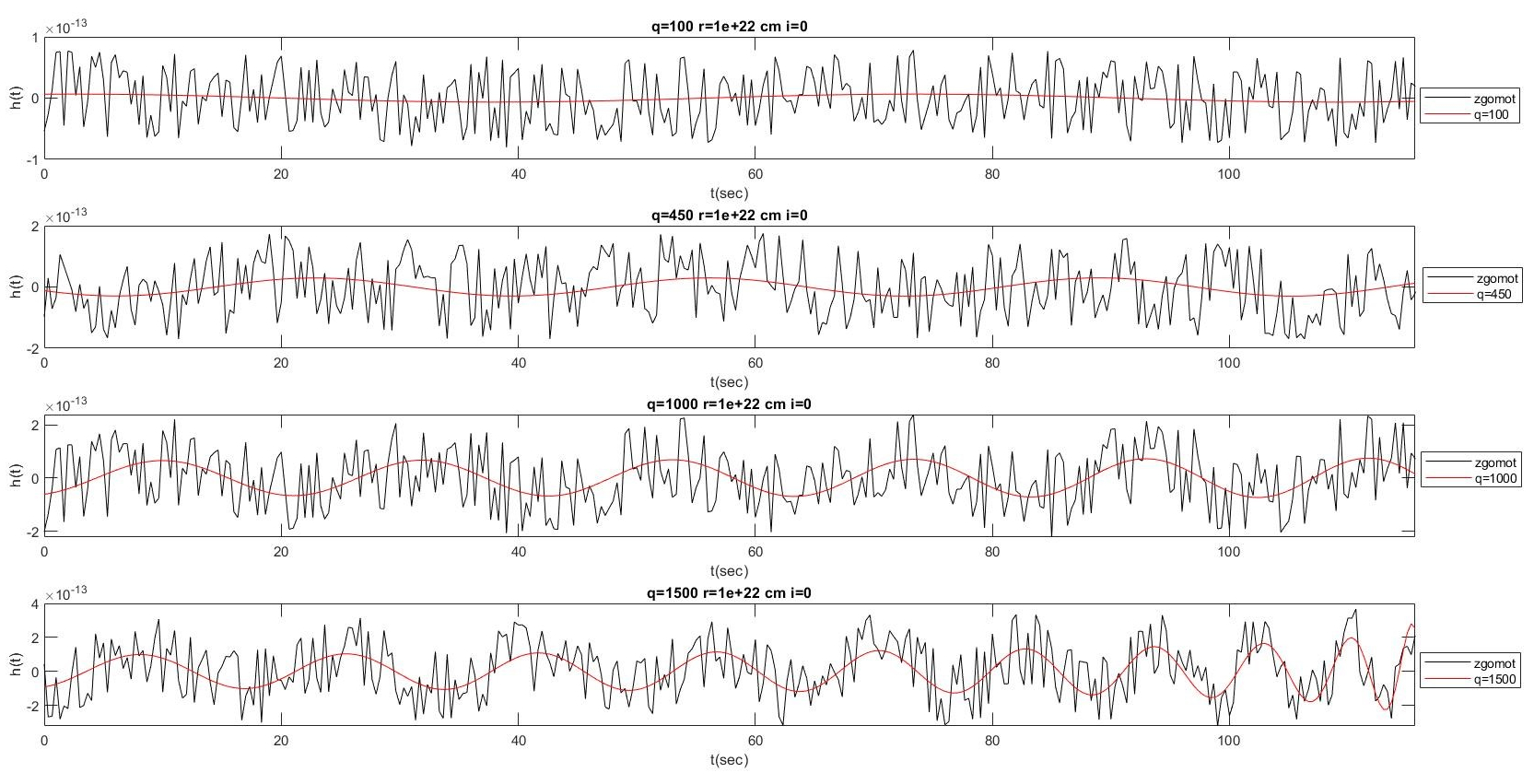}
\caption{Shape of the noise for four gravitational waves emitted by a binary system, for four different values of the source mass ratios.}
\label{fig:noisedifq}
\end{figure}

The gravitational wave simulator from \cite{Popescu2020} also has the option to add random noise to the gravitational wave signals. Figure \ref{fig:noisedifq} shows the waveforms together with the superimposed wave noise, for four different values of the mass ratio and a signal-to-noise ratio of $4.2$. The orbital inclination and distance to the source remain constant.

\begin{figure}[h!]
\centering
\includegraphics[viewport=0.5cm 0.3cm 20cm 7cm,clip,scale=0.8]{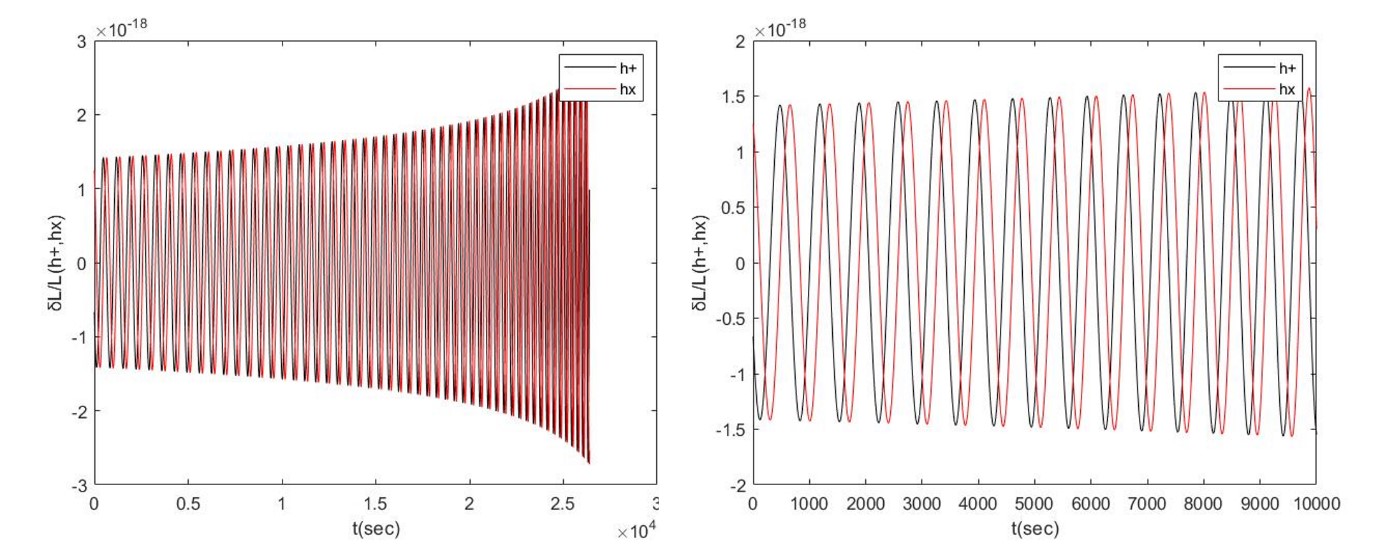}
\caption{The gravitational waves emitted by a binary system, separated into the two types of polarization, $h_{+}$ and $h_{x}$}
\label{fig:difpol}
\end{figure}

Figure \ref{fig:difpol} shows the superimposed polarization modes, $h_{+}(t)$ and $h_{x}(t)$, for the gravitational waves emitted by a binary system in which one of the objects has mass $m_1 = 9 \times 10^4 M_{\odot}$, $q = 12$. The right panel in Figure \ref{fig:difpol} represents a zoom of the left panel. The shift between the two polarization modes occurs because the polarization $h_{+}$ depends linearly on a $cos()$ function while $h_{x}$ depends linearly on a $sin()$ function. Since $cos(x) = sin(x) + \pi/2)$, the time shift between the two is given by the time needed by the system to swipe $\pi/2$ of a full orbit. But, as the system is closer to the collision, the orbital frequency increases and the trigonometric "distance" of pi / 2 is traveled faster and faster. 

The simulator was instructed to stop  at the innermost stable circular orbit, beyond which the quadrupolar approximation is no longer valid.
At this stage, the data used to train and test the LL BiLSTM network consists only of $h_{+}$ polarized waveforms. The parameter with respect to which the neural network performs the classification of the gravitational waves is the mass ratio of of the two objects in the binary system, \textit{q}, while the orbital inclination, \textit{i} and the distance to the source, \textit{r} are fixed to $i = 0 rad$ and $r = 10^{22} cm$.

\section{Results}
 
\subsection{Training and Testing Data Preparation}
\label{section:dataprep}
We have used two distinct data sets for training and testing our neural networks:
\begin{itemize}
    \item For the 2-class algorithm we have used a set of 201 data files, representing time variations of the gravitational wave amplitude over an interval of 16 minutes and with a time step of 0.01 seconds. Out of these, 101 were H-class data and 100 L-class data.
    \item For the 4-class algorithm we have used a more inhomogeneous data set, consisting of 100 class A data, 550 class B data, 601 class C data and 402 class D data. This time, neither the duration nor the time step of the time series are uniform, ranging from an interval of 4 to 16 minutes and time steps from 0.3333 to 0.01 seconds. Also, for class D, we have included in the analysis waveform data with superimposed random noise (see Figure \ref{fig:noisedifq}). Since the high degree of inhomogeneity in these data set is likely to affect the training process, it is necessary to further process them. First, we have segmented the signals into equal length sequences, each having 32000 points. This is an useful operation since during training, the data is automatically divided into mini-data packets which are then padded or truncated to have the same length. Too much truncation or padding can have a negative effect on the network performance, as the network may misinterpret a signal due to too much data being added or cut. That is why we have chosen to do the segmentation before training. For this, we have used the \textit{segmentSignals} function developed by Gari et al. (\cite{Gari2017}) and set the segment length to 32000.  The function ignores signals with less than 32000 lines and truncates longer signals into segments of 32000 length. The segment length was chosen such that the data loss was minimal. After the segmentation, we obtain a total number of 4067 "segments" (time series), out of which only about 7\% are from class A and 7\% from class D, whereas data from classes B and C represent 44\% and 40\% respectively. This can lead to biases in the classification since the neural network algorithm could learn that it can achieve high accuracy if it classifies all data as being either B or C. In order to avoid this bias, we have multiplied the number of class A and D files by "cloning" them with the \textit{repmat} function.
\end{itemize}

We have randomly divided the data into two subsets, one for training the neural network and one for testing classification accuracy. For both versions of the algorithm, we have used 90\% of the data for training and 10\% for testing.

\subsection{LL BiLSTM Network Configuration}

\begin{figure}[h!]
\centering
\includegraphics[scale=1]{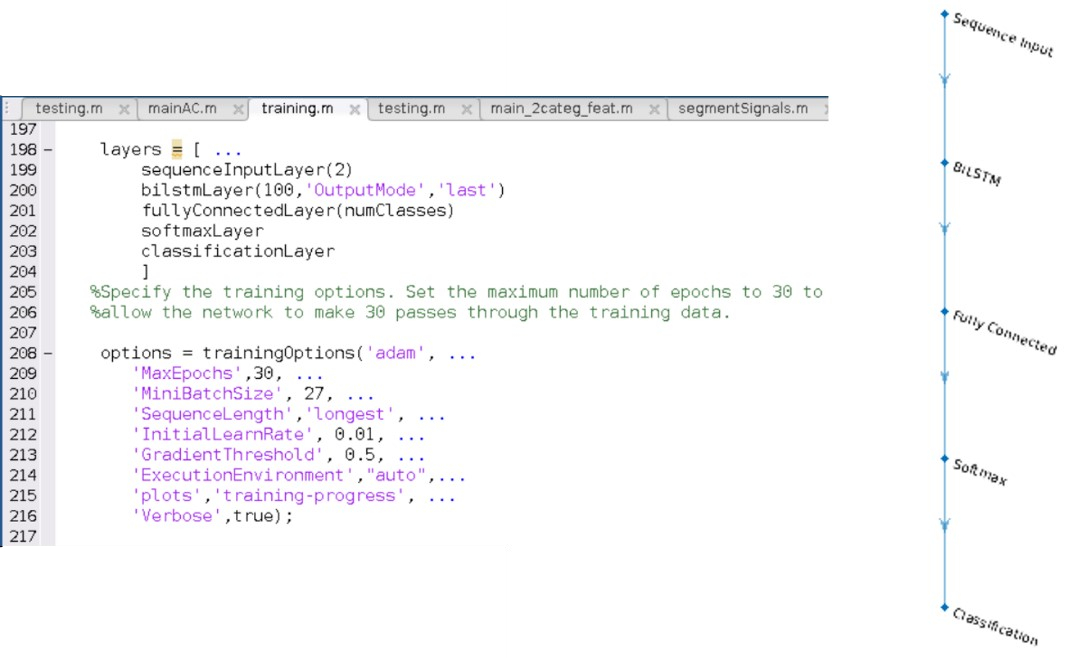}
\caption{Neural network architecture and features.}
\label{fig:fig30}
\end{figure}

As described in Section \ref{section:main}, LSTM deep neural networks can learn long-term time dependencies between sequential data steps in a time series. In addition, a bidirectional LSTM  layer swipes the time series in two directions, both forward and backward.
Figure \ref{fig:fig30} illustrates the architecture of our network, LL BiLSTM, as well as the options chosen for training.

The network has 5 layers:
\begin{itemize}
    \item A \textit{sequence input layer}. This layer feeds the training data into the network. In the final version of the algorithm, this layer has 2 dimensions, representing the two features
    \item A \textit{bidirectional LSTM layer}. This layer has an output of size 100. The bidirectional LSTM maps the input time series into 100 features and then prepares the output for the next layer.
    \item A \textit{fully connected layer}. Here we specify the total number of classes (2 or 4). The role of the fully connected layers is to connect every neuron from one layer with every neuron from another.
    \item A \textit{softmax layer}. The name of this layer comes from the \textit{softmax function}, a function that transform all the elements of a vector from real values into probabilities with values between 0 and 1. The input values can have any value.
    A softmax layer is useful for neural network classifiers, but only if the classes are mutually exclusive. It is usually placed at the end of a multilayered network, since it converts the real-value output scores, that may be difficult to display or use as input, to a normalized probability distribution.  
    \item A \textit{classification layer}. This layer computes cross-entropy loss in the case of multi-class classification problems where the classes are mutually exclusive. This layer infers information about the number of classes from the previous layers.
\end{itemize}
We have used the following training options:
\begin{itemize}
    \item An \textit{Adaptive Mode Estimation}(ADAM) solver. This type of solver works efficiently with neural networks classifiers.
    \item The maximum number of epochs (\textit{MaxEpochs} option), meaning the number of times the network passes through the training data, is sufficient for obtaining maximum accuracy and a good training time.
    \item The size of the mini-batch. As mentioned above, during training process the neural network divides the data into mini-batches. Then, the time series in each mini-batch are padded or truncated so that they are the same length. A size too small (\textit{MiniBatchSize} option) of the mini-batches may decrease the accuracy of the algorithm. On the other hand, a size too large could lead to memory overflow situations.
    \item The initial learning rate. The \textit{InitialLearnRate} option accelerates training. A too high initial learning rate can result in arbitrary spikes in the loss. On the other hand, a too low learning rates increases considerably the training time.
    \item The sequence lenght. The \textit{SequenceLength} option divides the signal into smaller parts so that the computer does not run out of memory by "looking" at too much data at once. On the other hand, a too short sequence length can lead to a wrong learning of the signal shape.
    \item The gradient threshold. The \textit{GradientThreshold} option helps stabilizing the training process, by preventing the gradients from becoming too large.
    \item The execution environment. The \textit{ExecutionEnvironment} option can be set to \textit{auto/cpu/gpu/multi-gpu/parallel}
    \item The \textit{plots} option is an option that determines which graphics to display during the training process.
    \item The \textit{verbose} option is a boolean option for switching on and off the text messages concerning the training progress.
\end{itemize}

\subsection{2-Class LL BiLSTM Network}
\begin{figure}[h!]
\centering
\includegraphics[viewport=0.1cm 0cm 15cm 15cm,clip,scale=0.8]{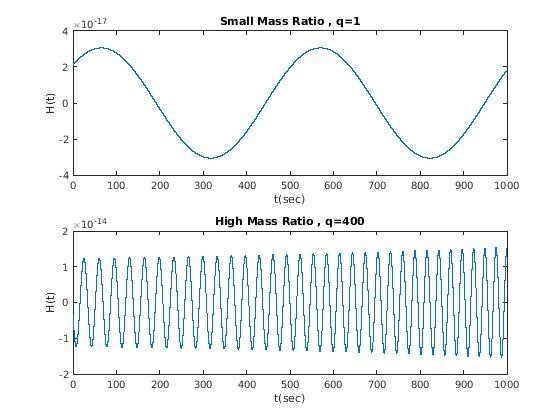}
\caption{Temporal evolution of the simulated GW amplitude, for a sample in each class (2-class algorithm).}
\label{fig:fig31}
\end{figure}

As mentioned above in section \ref{section:dataprep}, the first step of our analysis was to develop a BiLSTM neural algorithm that distinguishes between two classes of gravitational waveforms. We have used 101 class H (High, $q\in[400-500]$) data and 100 class L (Low, $q\in[1-10]$) data. Out of the total set of simulated data, we have randomly picked 91 for class H and 90 for class L for training and 10 of each class for testing (so 90\% training and 10\% testing).
In Figure \ref{fig:fig31} a sample from each class is plotted in order to easily visualize how the shape of the waveforms varies depending on different values of the binary source mass ratio.

\clearpage

\begin{figure}[h!]
\centering
\includegraphics[scale=1]{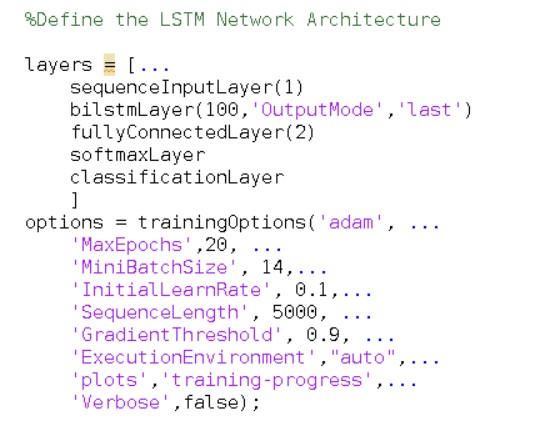}
\caption{Network architecture and options,when the actual values of the time series are used for training}
\label{fig:fig32}
\end{figure}

\begin{figure}[h!]
\centering
\includegraphics[scale=1]{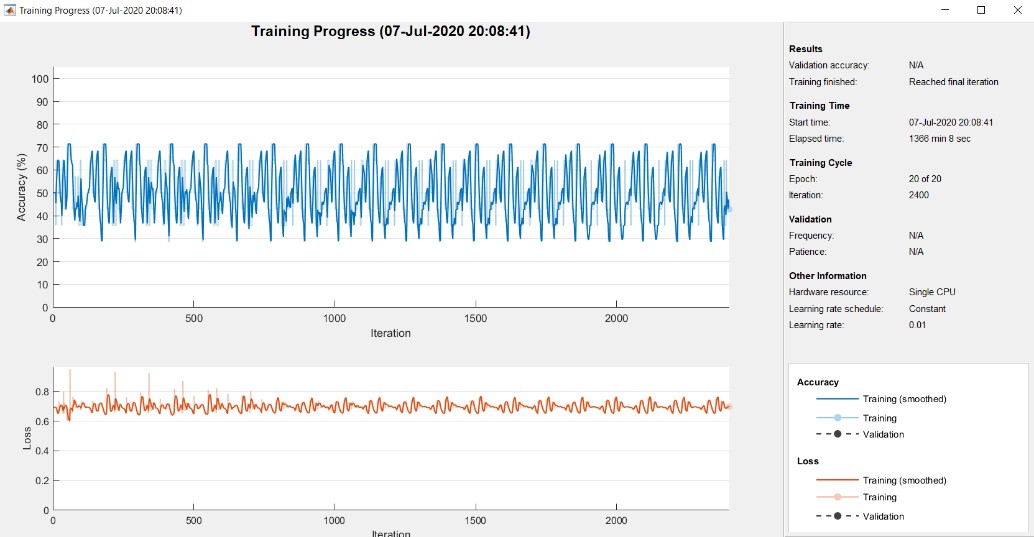}
\caption{The progress of the training process for LL BiLSTM network with 2 classes, when the actual values of the time series are used for training. Horizontal top panel: the evolution of the classification accuracy as a function of the number of iterations. Bottom horizontal panel: Variation of the loss function during the training process. Vertical panel on the right: information about the training process (status, training time, current era and iteration, etc.).}
\label{fig:fig33}
\end{figure}

\begin{figure}[h!]
\centering
\includegraphics[scale=0.6]{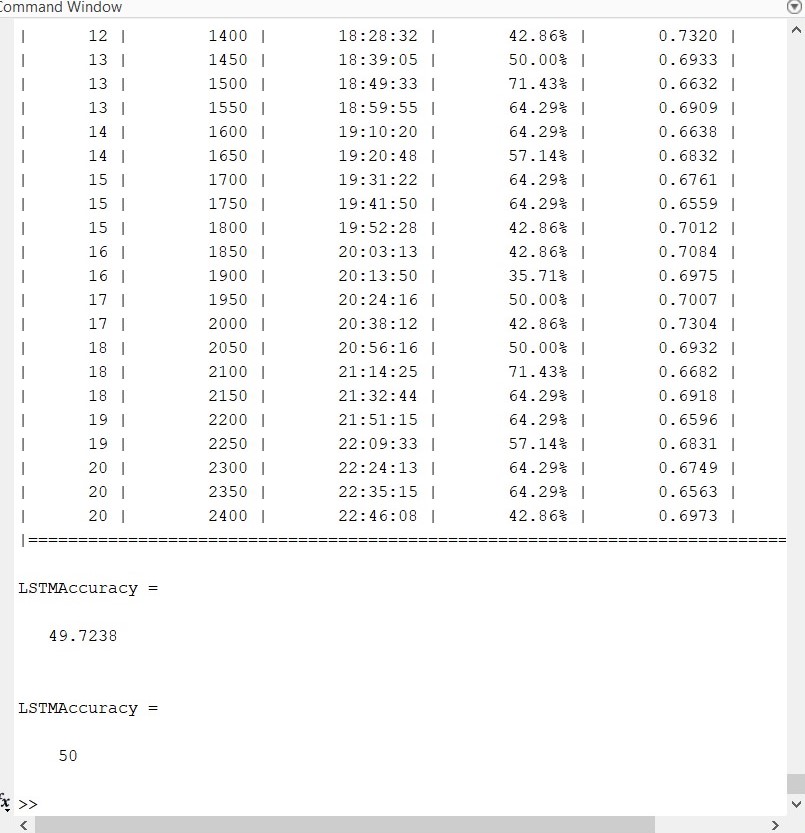}
\caption{Accuracy of the training process for the 2-class LL BiLSTM network, when the actual values of the time series are used for training}
\label{fig:fig34}
\end{figure}

At this early stage, we have tried training our network using the actual values of the time series. The network architecture and options are given in Figure \ref{fig:fig32}. Since the signals have one dimension each, we set the \textit{sequenceInputLayer} option to 1.
Figure \ref{fig:fig33} shows the progress of the training process. The upper horizontal panel shows the evolution of the classification accuracy as of function of the number of iterations, while the lower horizontal panel shows the variation of the loss during the training process. In the vertical panel on the right, you can see information about the training process (status, training time, current time and iteration, etc.).
This first training attempt did not have very good results. It can be seen from Figures \ref{fig:fig33} and \ref{fig:fig34} how the accuracy stagnates around 60\%, the loss is high and the running time is very long.

\clearpage

\begin{figure}[h!]
\centering
\includegraphics[viewport=2cm 0cm 20cm 15cm,clip,scale=0.8]{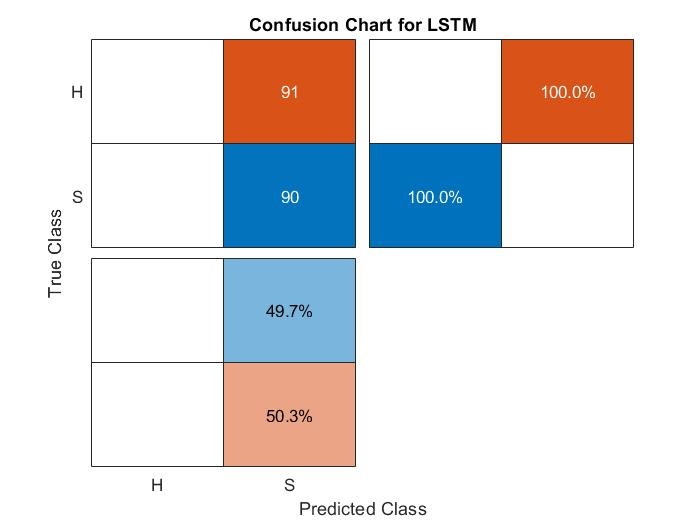}
\caption{The confusion matrix of the 2-class LL BiLSTM network (2-class LL BiLSTM), obtained at the end of the training.}
\label{fig:fig35}
\end{figure}

\begin{figure}[h!]
\centering
\includegraphics[viewport=2cm 0cm 20cm 15cm,clip,scale=0.8]{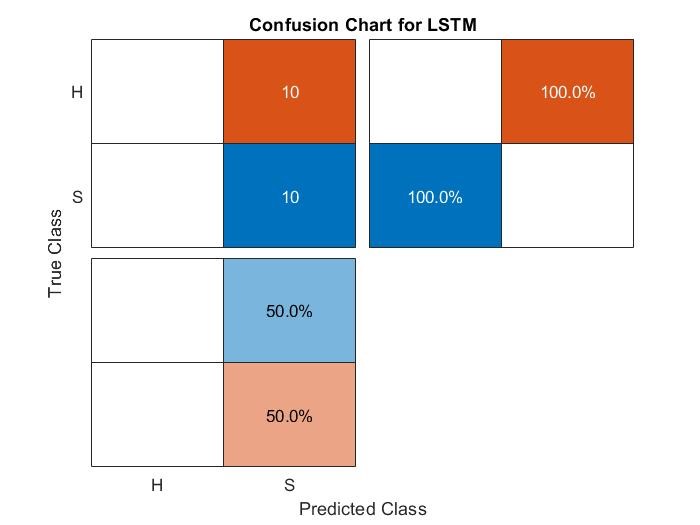}
\caption{The confusion matrix of the 2-class LL BiLSTM network (2-class LL BiLSTM), obtained at the end of the testing.}
\label{fig:fig36}
\end{figure}

Also, by looking at both confusion matrices, the training one (Figure \ref{fig:fig35}) and the testing one (Figure \ref{fig:fig36}), it can be seen that none of the training or testing data from class H were classified correctly. The algorithm classified all data as class S.

\clearpage

In order to improve the overall performance of the algorithm we have used feature extraction techniques based on fast Fourier transform analysis, which is a common optimization method employed in machine learning. Here, we use time-frequency moments to extract information from the spectrograms, each moment being used as a one-dimensional feature used as input for our BiLSTM networks:
\begin{itemize}
    \item \textit{Instantaneous frequency}, a function that estimates the time-dependent frequency of a signal as the first moment of the power spectrogram, and computes a spectrogram using short-time Fourier transforms over time windows. The time-dependent outputs of the function are computed at the centers of the time windows.
    \item \textit{Spectral entropy}, a function that measures how flat or spiky is the spectrum of a signal. It is also computed based on the power spectrogram and the time-dependent output values correspond to the centers of the time windows.
\end{itemize}

\begin{figure}[h!]
\centering
\includegraphics[viewport=0.3cm 0cm 20cm 12cm,clip,scale=0.9]{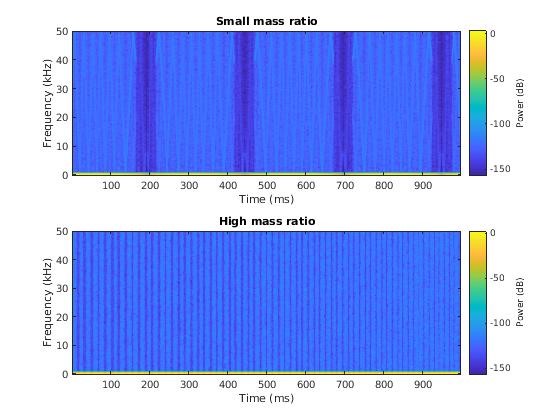}
\caption{Spectrograms of the simulated GW, a sample from each class (2-class algorithm).}
\label{fig:fig37}
\end{figure}

\begin{figure}[h!]
\centering
\includegraphics[viewport=0.3cm 0cm 20cm 12cm,clip,scale=0.9]{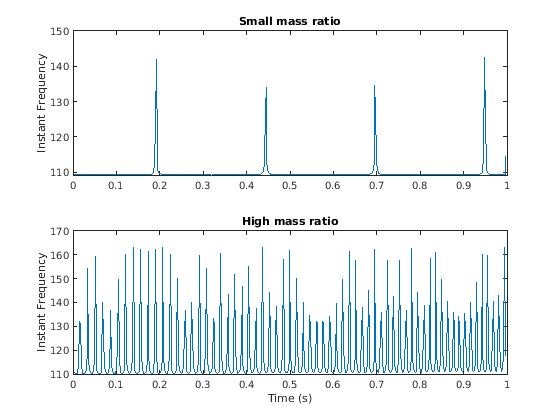}
\caption{Instantaneous frequency of the simulated GW, a sample from each class (2-class algorithm).}
\label{fig:fig38}
\end{figure}

\begin{figure}[h!]
\centering
\includegraphics[viewport=0.3cm 0cm 20cm 12cm,clip,scale=0.9]{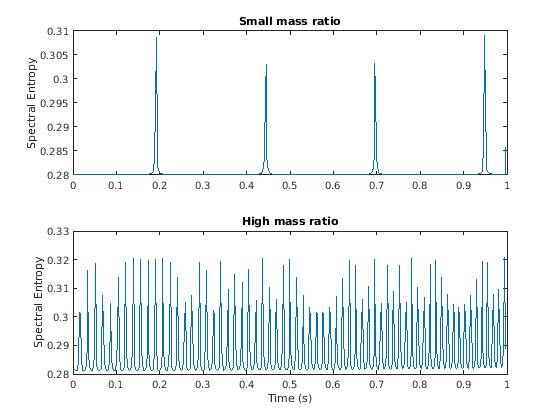}
\caption{Spectral entropy of the simulated GW, a sample from each class (2-class algorithm).}
\label{fig:fig39}
\end{figure}

This type of analysis requires the Matlab's \textit{Signal Processing Toolbox™}.
Figure \ref{fig:fig37} shows the spectrograms of two waveforms from the test data set, for the case of the 2-class LL BiLSTM network, one from each class. One can clearly see the differences between the spectrograms for each of the two classes. The same differences are visible in Figures \ref{fig:fig38} and \ref{fig:fig39}, which illustrate the instantaneous frequencies and spectral entropies of two waveforms in the test data set, for the case of the 2-class LL BiLSTM network, one from each class.

\clearpage

\begin{figure}[h!]
\centering
\includegraphics[viewport=0.3cm 0cm 20cm 12cm,clip,scale=0.7]{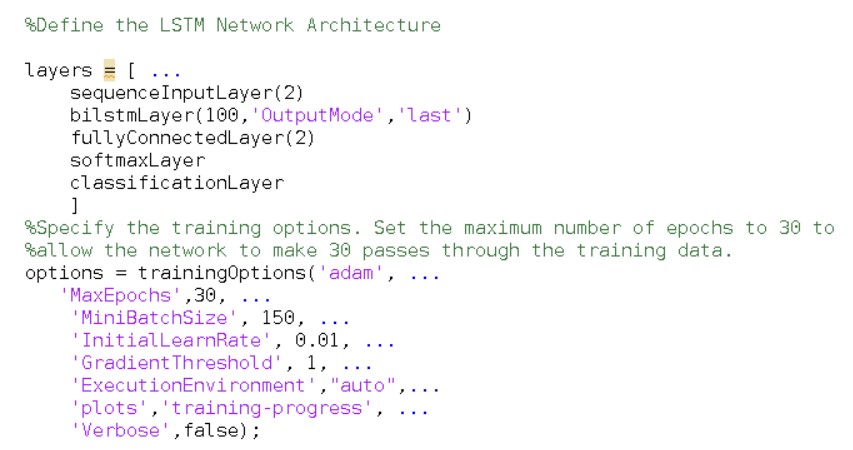}
\caption{Architecture and training options of the LL BiLSTM network with 2 classes, when the spectral features of the time series are used for training.}
\label{fig:fig40}
\end{figure}

The network architecture and training options for this version of the algorithm are given in Figure \ref{fig:fig40}. Because this time we assign two time-frequency moments to each signal, namely the instantaneous frequency and the spectral entropy, the \textit{sequenceInputLayer} option has two dimensions. We set the number of epochs to 30 and the size of mini-batches to 150, values estimated to be optimal to maximize network performance.

\clearpage

\begin{figure}[h!]
\centering
\includegraphics[viewport=0.3cm 1cm 20cm 8cm,clip,scale=0.9]{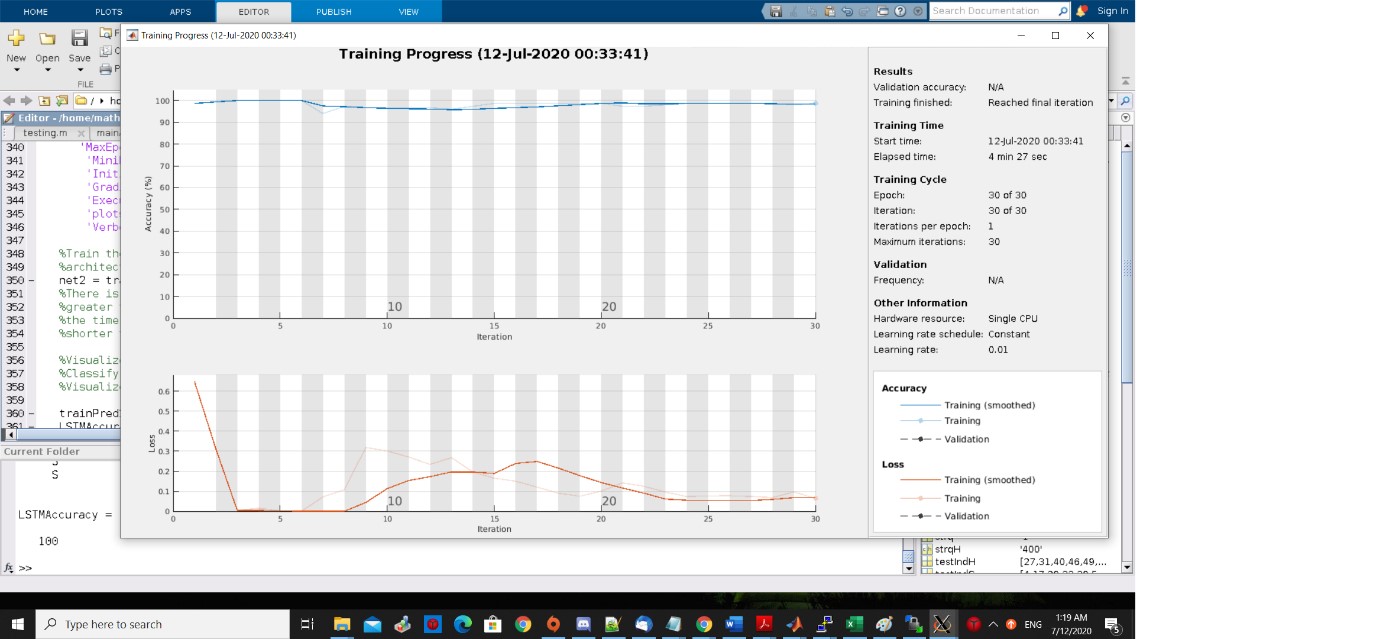}
\caption{Training progress of the LL BiLSTM network with 2 classes (2-class LL BiLSTM). The upper plot: the evolution of accuracy during the training process. The lower plot: the loss variation during the trainig process. Right-hand side panel: informations about the training process (status, training time, epoch and current iteration, etc.)}
\label{fig:fig41}
\end{figure}

The graph of the new training process, \ref{fig:fig41}, shows a considerable improvement in the training accuracy (which becomes 100\%) and the training time (which decreases from about 22 hours to about 4 minutes).

\clearpage

\begin{figure}[h!]
\centering
\includegraphics[viewport=0.3cm 0cm 14cm 12cm,clip,scale=0.8]{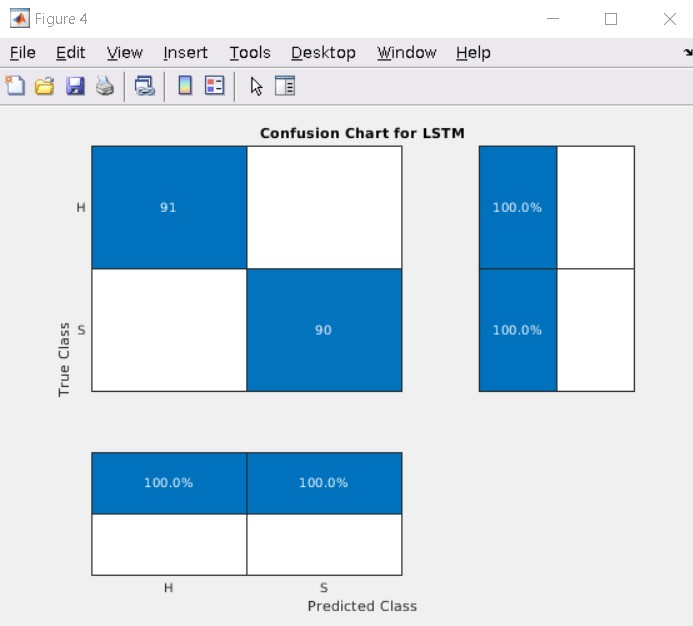}
\caption{The confusion matrix resulting from the training process of the LL BiLSTM network with 2 classes that uses the spectral characteristics.}
\label{fig:fig42}
\end{figure}

\begin{figure}[h!]
\centering
\includegraphics[viewport=0.3cm 0cm 14cm 12cm,clip,scale=0.8]{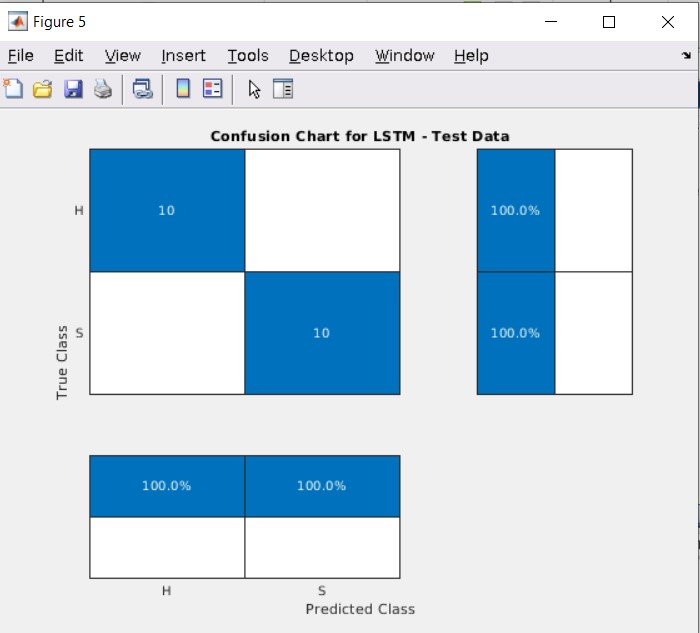}
\caption{The confusion matrix resulting from the testing process of the LL BiLSTM network with 2 classes that uses the spectral characteristics.}
\label{fig:fig43}
\end{figure}

It can also be seen, both from the training (Figure \ref{fig:fig42}) and testing (Figure \ref{fig:fig43}) confusion matrices that all training and testing data were classified correctly.

\clearpage

\subsection{4-Class LL BiLSTM Network}

\begin{figure}[h!]
\centering
\includegraphics[viewport=0.3cm 0cm 20cm 12cm,clip,scale=0.65]{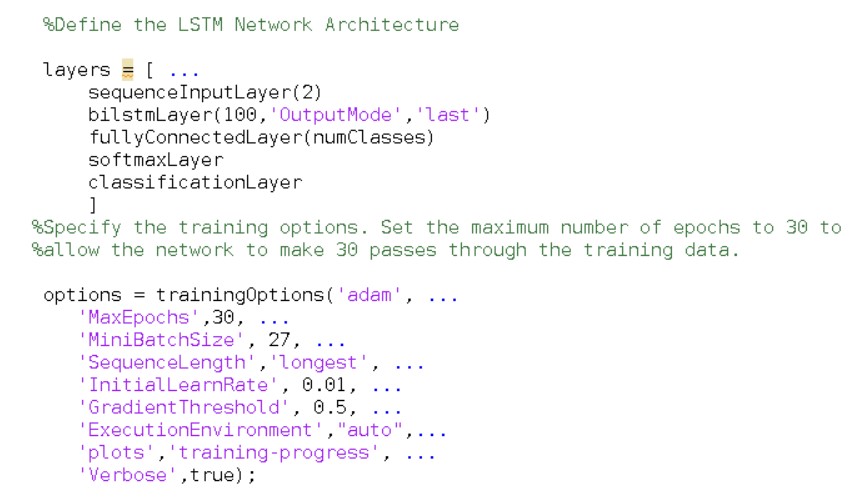}
\caption{Architecture and training options of the 4-class LL BiLSTM network.}
\label{fig:fig44}
\end{figure}

The next step in our analysis was to implement a more complex version of the BiLSTM neural network that classifies waveforms according to 4 classes, corresponding to 4 mass ratio intervals, A ($q\in[1-10]$), B ($q\in[300-350]$), C ($q\in[400-500]$) and D ($q\in[750-950]$). As we mentioned in \ref{subsection:trainandtestdata}, we used a much more complex data set, in which neither the duration nor the time step of the time series are uniform. Also, for class D, we included in the analysis data consisting of waveforms plus random noise.
The network architecture and training options for this version of the algorithm are given in Figure \ref{fig:fig44}. Because this time the classification is done with respect to 4 classes and the data set is larger and much more complex, we reduce the size of the mini-batches to 27, set the gradient threshold to 0.5, values that we have estimated to be optimal for maximizing the network performance.

\clearpage

\begin{figure}[h!]
\centering
\includegraphics[viewport=0.3cm 0cm 20cm 12cm,clip,scale=0.9]{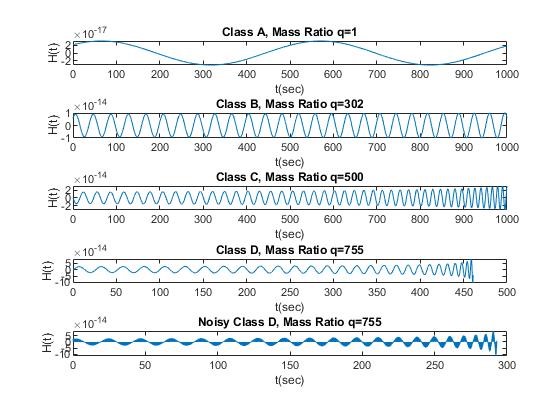}
\caption{Temporal evolution of the simulated GW amplitude, for a sample in each class (4-class algorithm). The fifth plot illustrates a sample of the random-noise perturbed class D subset data.}
\label{fig:fig45}
\end{figure}

\begin{figure}[h!]
\centering
\includegraphics[viewport=0.5cm 0cm 20cm 12cm,clip,scale=0.8]{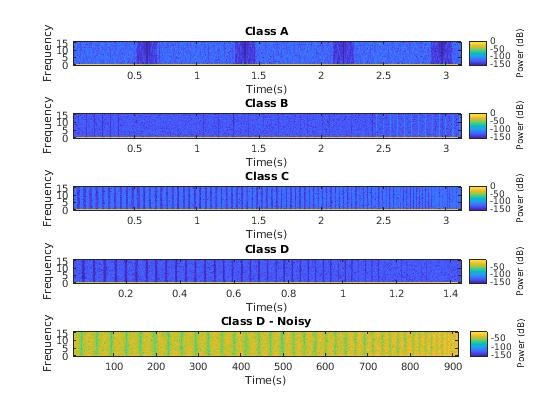}
\caption{Spectrograms of the simulated GW, a sample from each class (4-class algorithm). The fifth plot illustrates the spectrogram of a sample of the random-noise perturbed class D subset data.}
\label{fig:fig46}
\end{figure}

Just as in the case of the 2-class, from the total set of simulated data we have randomly chosen 90\% for training and 10\% for testing.
After the initial data processing, described in detail in \ref{subsection:trainandtestdata}, which was essential considering their complexity, we visualized the data and initiated the training.
In Figure \ref{fig:fig45}, you can see how the waveforms vary depending on different values of the source mass ratio. Also, in the last plot of  Figure \ref{fig:fig45} it can be seen how the waveform changes when random noise is added on top of it.
Figure \ref{fig:fig46} shows the spectrograms of four waveform samples from the training data set, one from each class, and another spectrogram (the bottom graph in the image), for the class D data subset with random noise. One can clearly see the differences between the five spectrograms corresponding to the differences in the source mass ratio. This can be also seen in Figures \ref{fig:fig48} and \ref{fig:fig49} in which we have plotted the instantaneous frequencies and spectral entropies for five waveforms, one sample from each of the A, B and C classes and two from class D, one with noise and one without.

\begin{figure}[h!]
\centering
\includegraphics[viewport=0.3cm 0cm 20cm 12cm,clip,scale=0.8]{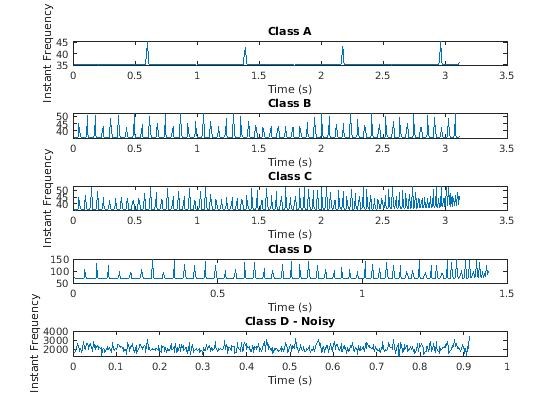}
\caption{Instantaneous frequency of the simulated GW, a sample from each class (4-class algorithm). The 5th plot represents one sample from the random-noise perturbed class D sub-set.}
\label{fig:fig47}
\end{figure}

\begin{figure}[h!]
\centering
\includegraphics[viewport=0.3cm 0cm 20cm 12cm,clip,scale=0.8]{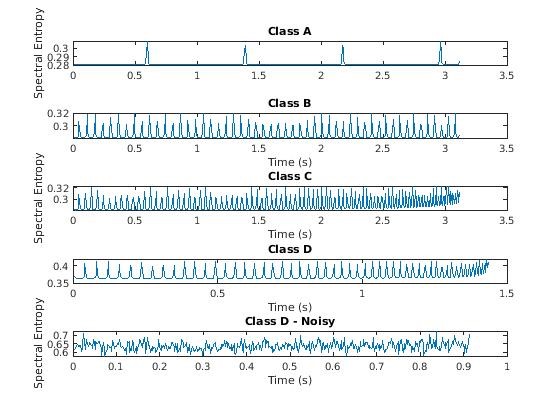}
\caption{Spectral entropy of the simulated GW, a sample from each class (4-class algorithm). The 5th plot represents one sample from the random-noise perturbed class D sub-set.}
\label{fig:fig48}
\end{figure}

\clearpage

As it can be seen from Figure \ref{fig:fig49} showing the training progress, the accuracy of the training is very good despite the fact that both the algorithm and the data are more complex. We have obtained an overall accuracy of 95.5741\% and a loss that is very close to 0.

\begin{figure}[h!]
\centering
\includegraphics[viewport=0.3cm 0cm 20cm 12cm,clip,scale=0.77]{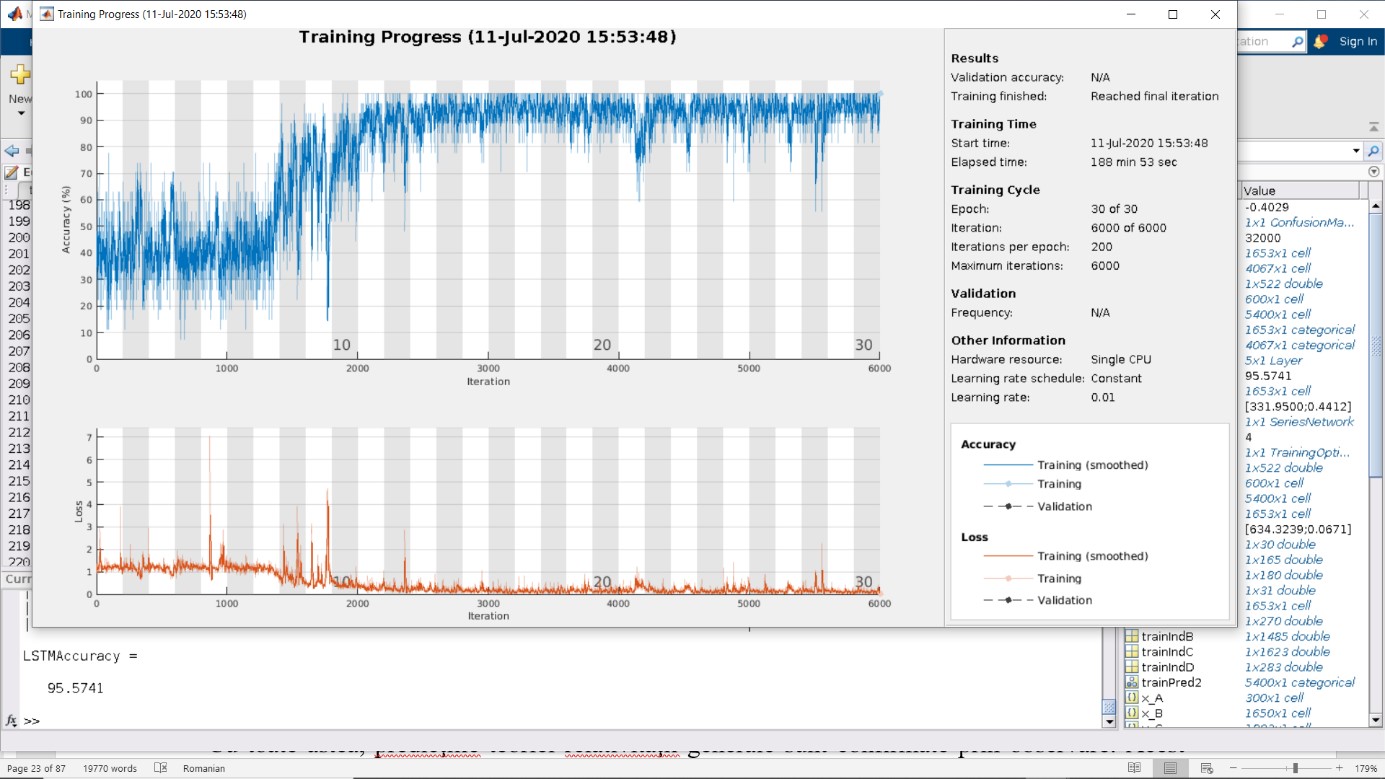}
\caption{Training progress window. The training can be stopped by pressing the round button with a black square in the middle, located in the bottom-right area of the progress window.}
\label{fig:fig49}
\end{figure}

\clearpage

\begin{figure}[h!]
\centering
\includegraphics[viewport=2cm 0cm 20cm 15cm,clip,scale=0.8]{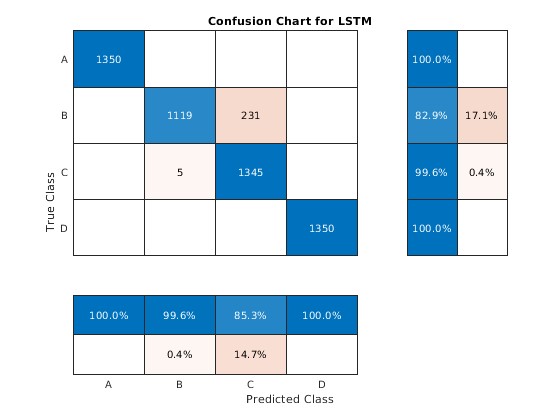}
\caption{The confusion matrix resulting from the training process of the 4-class LL BiLSTM network.}
\label{fig:fig50}
\end{figure}

\begin{figure}[h!]
\centering
\includegraphics[viewport=2cm 0cm 20cm 15cm,clip,scale=0.8]{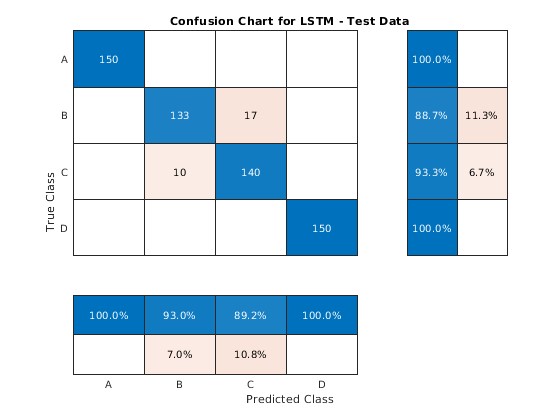}
\caption{The confusion matrix resulting from the testing process of the 4-class LL BiLSTM network.}
\label{fig:fig51}
\end{figure}

Also, analyzing both the training (Figure \ref{fig:fig50}) and the testing (Figure \ref{fig:fig51}) confusion matrices, we can see observed that all the training and testing data from classes A and D, and 99.6\% training/93\% testing from class B and 85.3\% training/89.2\% testing from class C were correctly classified. This result is extremely good, given the lack of homogeneity of the input data and the very close values of the mass ratio ranges for classes B and C.

\clearpage

\section{Conclusions}

In this paper, we presented the first results obtained in the development of a fast and accurate method to analyse signals coming from gravitational wave detectors, with applications in the development of low-latency alert generation systems for gravitational wave observatories (like the future LISA Mission).
The method is based on a neural network algorithm trained to recognize and characterize gravitational wave patterns in signal + noise data samples. 
Neural networks have already proven effective in several fields, including gravitational wave astronomy where they were used for the detection and characterization of gravitational wave signals emitted by binary black hole systems.
We implemented a neural network that we trained to recognize gravitational waveforms and classify them according to the parameters of the emitting source. The neural network was developed based on a Bidirectional Long-Short Term Memory (BiLSTM) deep learning algorithm. At this stage, the algorithm was trained to classify the gravitational waveforms emitted by binary systems of compact astronomical objects according to the mass ratio of the objects forming the system. For the training and testing of the network we used mock data representing gravitational wave signals (gravitational wave amplitude as a function of time) emitted by black hole binary systems, simulated using an in-house code \cite{Popescu2020}.
The algorithm was developed in Matlab and contains several modules, embedded in an application.
We started by creating a neural network based on a BiLSTM algorithm, LL BiLSTM, that discerns between gravitational waveforms from two distinct classes. Initially, we tried to train this network using the signal as it is, i.e. using as input values the actual time series. In this configuration, the training process led to very poor results both in terms of execution time and classification accuracy.
Following this first unsuccessful training attempt, we have decided to employ feature extraction techniques for optimizing the classifier, based on fast Fourier transform analysis. Thus, in a second step, instead of using the time series for training, the network was trained using two features of the waveform function, namely the instantaneous frequency and the spectral entropy. The training of the neural network based on these spectral characteristics has considerably improved both the execution time (from 22 hours to 4 minutes) and the classification accuracy (from 60\% to 100\%).
As a next step in our analysis, we implemented a more complex version of the BiLSTM neural network that classifies waveforms according to 4 classes, corresponding to 4 mass ratio intervals, and also included noisy data for one of the classes.
The results of the training process for this version of LL BiLSTM are extremely good (95.5741\% total accuracy), taking into account the lack of homogeneity of the input data and the very close values of the mass ratio ranges of two of the four classes.

We conclude that the current version of the LL BiLSTM algorithm demonstrates the ability of a well-configured and calibrated Bidirectional Long-Short Term Memory software to classify with very high accuracy and in an extremely short time gravitational wave signals, even when they are accompanied by noise. Moreover, the performance obtained with this algorithm qualifies it as a fast method of data analysis and can be used as a low-latency pipeline for gravitational wave observatories like the future LISA Mission.

\paragraph{Acknowledgement}This work was supported by the ESA PRODEX project DSLISALLP and by the LAPLAS VI project. The results reported here were obtained using the Scientific Computing Facilities of the Institute of Space Science.

\bibliography{LLBiLSTM}

\end{document}